\documentclass{elsart}
\usepackage{graphicx}
\usepackage{amssymb}
\begin{document}
\begin{frontmatter}

\title{Gravitation, C, P and T symmetries and the Second Law}
\author{ Gabriel CHARDIN}
\address{DSM/DAPNIA/SPP, CEA/Saclay,
F-91191 Gif-sur-Yvette Cedex, France}

\begin{abstract}
The intimate links between gravitation and the second law are summarized and two less known relations between gravity and thermodynamics are studied. Firstly, the information cost required to operate a Maxwell's demon on a curved spacetime is estimated using the Kolmogorov-Sinai entropy. More importantly, the charge and time (C and T) reversal properties of the Kerr-Newman solution in General relativity show that this solution, similarly to the Dirac equation, appears to represent both a particle and its antiparticle and suggests a definition of antimatter in general relativity. This definition leads to a parameter free explanation of the cosmological constant term observed in the supernovae data. The relation of this definition of antimatter with the coupled systems through opposite time arrows studied by Schulman is also emphasized.
\end{abstract}
\end{frontmatter}

\section{Introduction : motivations}
The links between gravitation and thermodynamics are fundamental and exemplified in the four laws of black hole thermodynamics, first demonstrated by Bardeen, Carter and Hawking \cite{BCH}. However, these authors initially considered this analogy with thermodynamics as being purely formal. A fundamental idea was then proposed by Jacob Bekenstein \cite{bekenstein} in 1972, when he claimed that the formal identity between the area of a black hole and entropy was really a physical identity.

Bekenstein was strongly criticized by Carter and Hawking at the 1972 Les Houches summer school, where he had first presented his conjecture but, ironically, his expression for the entropy and the thermal emission of black holes was two years later demonstrated using a semiclassical calculation by Stephen Hawking himself \cite{hawking}.
More recently, Strominger and Vafa \cite{SV}, by counting the quantum microstates of a macroscopic black hole, have verified the Bekenstein-Hawking entropy formula, and several derivations \cite{carlip} further confirmed its universality, which appears independent of the particular physical string theory used to derive it.

Let us recall the main relations for this entropy and radiation mechanism. For a (non rotating and uncharged) black hole of mass M, the internal energy is $U = Mc^{2}$, and its entropy
$$S_{BH} = S_{Black Hole} = S_{Bekenstein-Hawking}= 4\pi M^{2}$$
when M is expressed in units of Planck mass. From this, we can derive the expression for the black hole temperature:
$$(k_{B}T)^{-1}=\frac{\partial S}{\partial U}=\frac{8\pi GM}{\hbar c^{3}}$$

The evaporation time, for a black hole of mass M is $\sim M^{3}$, with M again expressed in Planck mass units.
Note that the Bekenstein-Hawking entropy expression strongly suggests that spacetime is discrete at the Planck length scale:
$$L_{Planck}=\sqrt{\frac{\hbar G}{c^{3}}} \sim 10^{-33} cm$$
\section{ A Maxwell's demon on curved spacetime}
The motivation of this study was the argument, discussed initially with John Bell \cite{bell} and M.M. Nieto \cite{nieto}, that Morrison's antigravity \cite{morrison} ---antiparticles \textquotedblleft falling up"--- would a priori entail a vacuum instability. But this is not necessarily a problem in itself. In fact, as just noted, in the presence of a black hole, the vacuum {\it is} unstable.
The question then becomes: what amount of antigravity $\Delta g$ between particles and antiparticles will mimic Hawking's radiation, which should certainly be considered as an acceptable vacuum instability? With surprise, it was found \cite{chardin92,chardin93} that Morrison's antigravity, $\Delta g \sim 2g$, just leads to the expression of Hawking's temperature. But is antigravity compatible with the Second Law ? In particular, would not antigravity allow a Maxwell's demon to decrease the entropy of an isolated system by hiding photons in the vacuum ?

\begin{figure}[hbtp]
\begin{center}
\includegraphics[width=.7\textwidth]{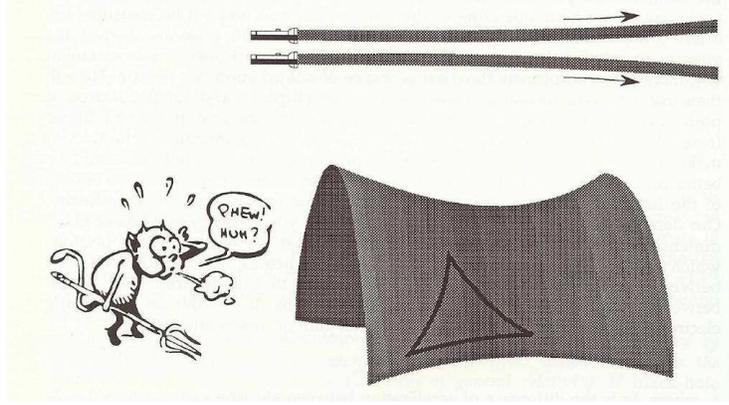}
\caption{ A Maxwell demon trying to violate the Second Law of Thermodynamics by using Morrison's antigravity, a violation of the Equivalence Principle, in order to hide photons in the \textquotedblleft vacuum" must pay an irreducible information cost due to the divergence of photon trajectories induced by the curvature of spacetime. This information cost is always larger than the expected gain of his manipulation}
\label{demon}
\end{center}
\end{figure}

However, operating a Maxwell's demon on curved spacetime has an information cost. If a demon wants to follow continuously particles or photons on a curved manifold, he must pay the price of the dynamical Kolmogorov-Sinai entropy, which can be estimated using Pesin's formula \cite{pesin}. The characteristic distance for the divergence of the photons is obtained from the components of the Riemann tensor, homogeneous to the inverse of the square of a curvature radius. This leads to a characteristic distance $\Delta z_{Lyap}$ in reference to the Lyapunov exponent :
$$\Delta z_{Lyap} \sim (\frac{GM}{c^{2}r^{3}})^{-1/2}$$

when the demon is operating at a distance r from a massive body of mass M.
In order to have a reasonable chance to suppress a photon in the box, of size at least $\Delta z_{Heisenberg}$ where he is operating, our demon will have to pay a price, in bits :
$$\Delta z_{Heisenberg} \frac{(\Delta z_{Heisenberg})^{3}}{( \alpha r_{C})^{3}} \frac{1}{\Delta z_{Lyapunov}}= \frac{r^{5/2}}{r_{C}r_{S}^{3/2}\alpha^{3}} > 1$$

where $\alpha$ is the fine structure constant and $r_{C}$ is the Compton radius of the electron. The first term is the minimal size of the enclosure where the demon is working, the second term is the number of collisions needed to absorb with a reasonable chance a photon in the fundamental state of the box, and the third term is the Lyapounov length, after which the demon must pay a price of one bit.
Obviously, the demon must realize his experiment at a distance of the center of attraction larger than the Schwarzschild radius $r_{S}$, and larger than the Compton wavelength of the electron $r_{C}$. Consequently, the demon is unable to operate efficiently enough to violate the Second Law of Thermodynamics \cite{chardin92,chardin93}.
\section{A Kerr-Newman electron is also a positron}
The motivation for reconsidering Morrison's antigravity was the coincidence, noted by Overhauser and Chardin \cite{chardin92,overhauser}, and known by Bell, that antigravity would explain CP violation in the neutral kaon system \cite{cpviolation}. But how could we justify antigravity ? To answer this question, let us study the Kerr-Newman solution, discovered by Roy Kerr \cite{kerr} and Ezra Newman \cite{newman}, and incorporating three parameters : mass $m$, specific angular momentum $a = L/m$, and charge $e$. We will summarize briefly the properties of this solution, assuming a fast Kerr solution, i.e. $e^{2} + a^{2} > m^{2}$, a condition met by all elementary particles, with the notable exception of the neutral Higgs boson \cite{visser}.

The fast Kerr geometry is particularly simple since it involves no horizon. The angular momentum imposes an annular shape to the singularity, which appears naked but nevertheless almost invisible since the measure of initial conditions allowing to reach the ring singularity is zero.
Brandon Carter has studied the topology of this solution in the late sixties \cite{carter}, noting the striking analogy that a \textquotedblleft Kerr-Newman electron" bears with real electrons. In particular, the gyromagnetic factor of the Kerr-Newman electron is $g = 2$ and the geometric extension of the ring is of the order of the Compton wavelength of the electron, giving it a spatial extension compatible with its cross-section.

Another interesting feature concerns the charge conjugation (C) properties of this solution. By crossing the interior of the ring, an observer will measure the charge and mass of the electron with a reversed sign. For a particle physicist, this means that if we were initially looking at an electron, we can now see a positron by crossing the interior of the ring. Most surprisingly, this \textquotedblleft positron" has a repulsive gravity. This results from the symmetry properties of the metric and electromagnetic field tensor form of the Kerr-Newman solution, which can expressed \cite{carter} in Boyer-Lindquist coordinates:
$$ds^{2}=-\frac{\Delta}{\rho^{2}}( dt-a \sin^{2}\theta d\phi)^{2}+\frac{\sin^{2}(\theta)}{\rho^{2}}\lbrack (r^{2}+a{2}) d\phi-a dt\rbrack^{2}+\frac{\rho^{2}}{\Delta}dr^{2}+\rho^{2} d\theta^{2}$$
and :
$$F=e\rho^{-4}\lbrack (r^{2}-a^{2}cos^{2}\theta )dr\wedge ( dt - a sin^{2} \theta d \phi ) +ar sin 2\theta d\theta \wedge \{ ( r^{2} + a^{2} ) d \phi - a dt\})$$
and where :
$$\Delta= r^{2} - 2Mr + a^{2} + e^{2}$$
and :
$$\rho^{2}=r^{2} + a^{2} cos^{2} \theta$$

Another significant and surprising feature of the Kerr-Newman solution is the fact that it is possible to go backward in time by exploring the negative mass part $(r < 0)$ of the solution. This feature was also studied by Brandon Carter \cite{carter}, and is known as the \textquotedblleft Carter time machine" \cite{oneill}. Considered initially as an indication of inconsistency, this can now be considered as an incentive to consider the possibility that antimatter, i.e. matter going backwards in time, can be defined in General Relativity.

\begin{figure}[hbtp]
\begin{center}
\includegraphics[width=.6\textwidth]{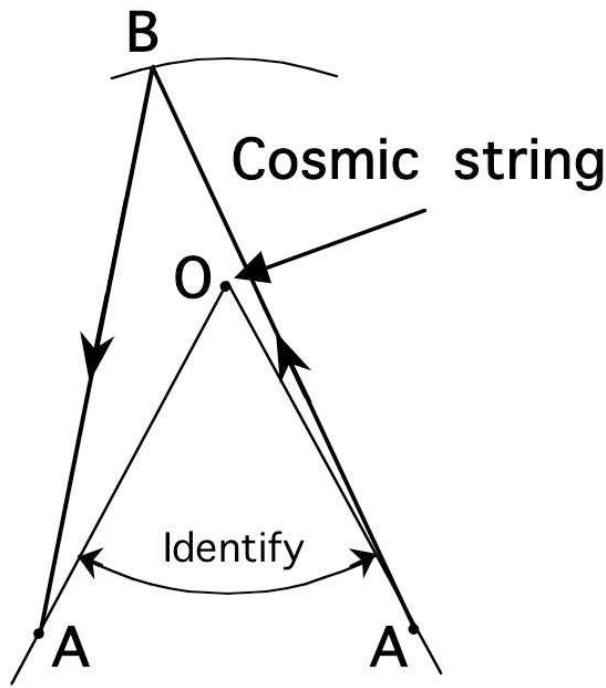}
\caption{ An observer in A can use a spinning cosmic string to discuss at zero time delay with a set of points B located on a portion of ellipsoid. In the special case where the angle deficit created by a cosmic string is $\pi$, a point B exists such that a signal emitted by A and reemitted by B comes back to A at zero time delay {\it and} with a direction identical to that of the initial signal. The interaction between A and B is then diverging, and B appears to be at the position of A}
\label{conjugate}
\end{center}
\end{figure}

\section{ Conjugate points in the Kerr-Newman geometry }
An important generic property of gravitation is the existence of points of infinite magnification for the image of an object through the \textquotedblleft lensing" created by a massive object. Used in recent years to detect massive compact halo objects (MACHOs) in our galactic neighborhood \cite{microlensing}, this magnification, when it is infinite, has the consequence that the lensed object may appear infinitely more luminous and closer than its true position. This property is even stronger for the fast Kerr-Newman geometry, where Closed Timelike Curves (CTCs) exist between any two points. For a given point A in the neighborhood of the ring, there exists a set of points B such that the radar interaction between A and B ---photons are emitted by A, scattered by B and received back by A--- is instantaneous. The signal emitted comes back with zero time delay as seen by the emitter, and the B points will give an observer in A the impression that they are sharing the same position.

These points can be explicitly constructed in the Kerr geometry in $2 + 1$ dimensions, where the spinning cosmic string is an exactly soluble model \cite{deser}. It is straightforward to demonstrate that the set of such points B lie on a portion of ellipse (Fig. 2) since the time pitch associated with a $2\pi$ rotation around the spinning cosmic string can be written as $\Delta t = 8 \pi aG$, where {\it a} is the specific angular momentum per unit length of the string.

From the existence of conjugate points in $2 + 1$ gravity, expected to be valid also in $3 + 1$ gravity from the existence of CTCs, there follows a definition of antiparticles in general relativity as the time-reversed {\it image} of particles observed through a Kerr ring. These Kerr rings are probably present in all elementary particles, if they are string loops, and in the past singularity of the Big Bang. This (non-local) definition of antimatter has the consequence that there exists a gravitational repulsion between matter and antimatter, defined relatively to each other and not in an absolute way. The coupling of systems with opposite arrows of time is reminiscent of the dynamical systems studied by Schulman \cite{schulman}. From the persistence of individual arrows of time for such weakly coupled systems, interactions of each system with the conjugate system are expected to appear as noise.

\section{ Apparent cosmological constant}
During the sixties and early seventies, several attempts have been made, using notably a conjectured repulsion in strong interactions \cite{omnes}, which failed to justify the survival of significant matter and antimatter domains in a symmetric matter-antimatter universe. The cosmological consistency of the previous definition of antimatter in general relativity has yet to be demonstrated. On the other hand, it should be noted that the usual arguments invoked to exclude the existence of large domains of antimatter through the non observation of diffuse gamma-ray background \cite{cohen} are not applicable in the case where diffusion and annihilation at the border of matter and antimatter domains is prevented by gravitational repulsion. Also, it is fascinating to note that the gravitational repulsion between matter and antimatter appears to lead to a natural explanation of the value of the cosmological constant observed in recent supernovae and CMB observations \cite{perlmutter,riess} :
$$\Omega_{tot}=\Omega_{baryon}+\Omega_{dark}+\Omega_{\Lambda}= 1 \pm 0.02$$
where a nearly flat universe is composed of only $\Omega_{baryon} \sim 4.5\%$ of ordinary matter, with a dark matter density $\Omega_{dark} \sim 0.3$ and an apparent cosmological density $\Omega_{\Lambda} \sim 0.7$.
To justify this statement, let us consider the expression for the deceleration parameter {\it q} as a function of the scale factor {\it a} and its derivatives:
$$q \equiv -\frac{a\ddot{a}}{\dot{a}^{2}}$$
When observed on a scale larger than the matter or antimatter domains, this symmetric universe will appear flat with a parameter $q <\sim 0$ due to the repulsion of adjacent domains. Fundamentally, there exists no cosmological constant, but if we insist to parametrize the repulsive term by a cosmological constant, this then implies:
$$q = \Omega_{matter}/2 - \Omega_{\Lambda} <\sim 0 \rightarrow \Omega_{\Lambda} = \mathcal{O}(1) \Omega_{matter}$$
Although, at any given epoch, this equality will locally be verified, the evolution of $\Omega_{matter}$ with time has for consequence that the derived value of the effective \textquotedblleft cosmological constant" will vary according to $( 1+z )^{3}$ since the matter density varies approximately in this way after recombination \cite{abl_chardin}. Therefore, there is no coincidence, only an (incorrect) assumption about the existence of a cosmological constant. Ripalda, in a different theoretical context, has also noted that repulsive gravity would lead to a cosmological constant of the correct order \cite{ripalda}.
\section{Conclusions}
We have summarized the relations between gravitation and thermodynamics, which appear extremely strong and universal. The Bekenstein-Hawking entropy relation, valid for a large class of string theories, can be considered as an indication that spacetime is discrete at the Planck length scale and that gravitation is the master arrow of time asymmetry.
In addition, charge (C) and time-reversal (T) properties of the Kerr-Newman solutions suggest a natural definition of antimatter in General Relativity. This definition provides a parameter free explanation of the otherwise extraordinary coincidence of the cosmological constant energy density with the matter density observed in the supernovae SN1a and CMB data. Defined as the time-reversed image of matter through Kerr wormholes, antimatter then provides with matter an explicit example of coupled systems with opposite arrows of time, as studied by Schulman.

\section{Acknowledgments}
Discussions with J-M. Rax, B. Carter, D. Connetable, G. Esposito-Far{\`e}se, E. Fischbach, T. Goldman, M.M. Nieto, R. Pain and A. Tzella are gratefully acknowledged. Needless to say, these people are not responsible for the errors and omissions contained in this paper.

\end{document}